\documentclass[
twocolumn,
superscriptaddress,
amsmath,amssymb,
aps,
prl,
]{revtex4-2}

\usepackage[T1]{fontenc}

\usepackage[english]{babel}
\makeatletter
\providecommand{\l@en}{\l@english}
\makeatother

\usepackage{physics,xfrac}

\usepackage{color}
\providecommand{\alert}[1]{{\color{red}\textbf{#1}}}

\usepackage{comment}
\usepackage{hyperref}

\usepackage[nameinlink]{cleveref}

\usepackage{xr-hyper}

\begin{document}

\title{Collective inhibition of light scattering from atoms into an optical cavity at a magic frequency}

\author{Á. Kurkó}
\email{curko.arpad@wigner.hun-ren.hu}
\affiliation{HUN-REN Wigner RCP, H-1525 Budapest P.O. Box 49, Hungary}
\author{B. Gábor}
\affiliation{HUN-REN Wigner RCP, H-1525 Budapest P.O. Box 49, Hungary}
\affiliation{Department of Theoretical Physics, University of Szeged, Tisza Lajos k\"{o}r\'{u}t 84, H-6720 Szeged, Hungary}
\author{D. Varga}
\affiliation{HUN-REN Wigner RCP, H-1525 Budapest P.O. Box 49, Hungary}
\affiliation{Department of Physics of Complex Systems, ELTE Eötvös Loránd University, Pázmány Péter sétány 1/A, H-1117 Budapest, Hungary}
\author{A.~Simon}
\affiliation{HUN-REN Wigner RCP, H-1525 Budapest P.O. Box 49, Hungary}
\affiliation{Department of Physics of Complex Systems, ELTE Eötvös Loránd University, Pázmány Péter sétány 1/A, H-1117 Budapest, Hungary}
\author{T. Barmashova}
\affiliation{HUN-REN Wigner RCP, H-1525 Budapest P.O. Box 49, Hungary}
\author{A. Dombi}
\affiliation{HUN-REN Wigner RCP, H-1525 Budapest P.O. Box 49, Hungary}
\author{T. W. Clark}
\affiliation{HUN-REN Wigner RCP, H-1525 Budapest P.O. Box 49, Hungary}
\author{F. I. B. Williams}
\affiliation{HUN-REN Wigner RCP, H-1525 Budapest P.O. Box 49, Hungary}
\author{D. Nagy}
\affiliation{HUN-REN Wigner RCP, H-1525 Budapest P.O. Box 49, Hungary}
\author{A. Vukics}
\affiliation{HUN-REN Wigner RCP, H-1525 Budapest P.O. Box 49, Hungary}
\author{P. Domokos}
\affiliation{HUN-REN Wigner RCP, H-1525 Budapest P.O. Box 49, Hungary}

\begin{abstract} 
We report on the observation of a new magic frequency within the hyperfine structure of the D2 line of ${}^{87}$Rb atoms at which the scattering of light into a high-finesse cavity is suppressed by an interplay between quantum interference and the strong collective coupling of atoms to the cavity. Scattering from a cloud of laser-driven cold atoms into the cavity was measured in a polarization sensitive way. We have found that both the Rayleigh and Raman scattering processes into the near-resonant cavity modes are extinguished at 185 MHz below the F=2$\leftrightarrow$F'=3 transition frequency. This coincidence together with the shape of the observed spectral dip imply that the effect relies on a quantum interference in the polariton excitations of the strongly coupled combined atom-photon system. We have also demonstrated the existence of a magic frequency around -506 MHz, where only the Raman scattering is suppressed due to a quantum interference effect at the single-atom level.
\end{abstract}

\date{\today}

\maketitle


\section{Introduction}

Interaction of atoms with the radiation field enclosed in a high-finesse optical resonator is the basis of many schemes aiming at the manipulation of atoms and photons at the quantum level. These include quantum information processing schemes \cite{reiserer_cavity-based_2015}, and extend as far as to superradiant lasers \cite{bohnet_steady-state_2012, norcia_superradiance_2016} and quantum phase transitions \cite{baumann_dicke_2010,klinder_observation_2015,landig_quantum_2016,leonard_supersolid_2017,kollar_supermode-density-wave-polariton_2017,mivehvar_cavity_2021,helson_density-wave_2023,ho_optomechanical_2025}. Photon scattering from laser-driven atoms into selected modes of a cavity is enhanced \cite{slama_cavity-enhanced_2007}, moreover, it can lead to very specific features for many-atom ensembles \cite{neuzner_interference_2016}. Even in the limit of coherent Rayleigh scattering,  
a high-level of complexity can emerge as a consequence of the spatial arrangement of atoms. Interference in the collective scattering can result in sub- and superradiance \cite{reimann_cavity-modified_2015,hotter_cavity_2023,yan_superradiant_2023}, or can yield phase transition like redistributions of the atomic ensemble \cite{domokos_collective_2002,black_observation_2003,arnold_self-organization_2012}. 

With drive laser frequencies closer to the atomic resonances, the photon scattering becomes non-trivial because of the multiplet structure of atomic hyperfine states \cite{pineiro_orioli_emergent_2022,veyron_effective_2022,suarez_collective_2023,hernandez_symmetry_2024}. The polarization of the incoming field can be rotated by Raman scattering processes in which the angular momentum change of light is compensated by that of the atomic ground state sublevels \cite{vrijsen_raman_2011,zhang_dicke-model_2018,davis_photon-mediated_2019}. Cavity enhancement of light scattering refers to Raman processes as well as to the Rayleigh scattering. The corresponding rates have, on the other hand, a strong dependence on the fine-tuned frequency across the range of hyperfine transitions. Interestingly, a magic frequency at -506 MHz from the $F=2 \leftrightarrow F'=3$ transition of the ${}^{87}$Rb D2 line was shown to lead to the suppression of Raman scattering due to a quantum interference effect occurring simultaneously for all the magnetic sublevels in the hyperfine ground state \cite{yan_superradiant_2023,masson_state-insensitive_2024}. We measured the scattering rates on an ensemble of atoms in the cavity in a broad frequency range, and confirmed this suppression effect at the given magic frequency.

The main result of this paper is that we found a new magic frequency which arises from the interplay between interference and \emph{collective effects} in the many-atom ensemble. We have observed a narrow dip at -185 MHz in the frequency-dependent rate of scattering into the cavity. In the regime of strong collective coupling of atoms to the cavity modes, excitations in the system can be generated in the form of polaritons. The coherent excitation of such polaritons is inhibited by a destructive interference in the collective polarizability of multilevel atoms at -185 MHz. This magic frequency is determined by the atomic properties and not by that of the cavity. Since the inhibition relies on a collective effect, the magic frequency is largely insensitive to inhomogeneities of uncontrolled fields within the atom cloud.

\section{Light scattering into the cavity}

\begin{figure}[htbp]
\includegraphics[width=\linewidth]{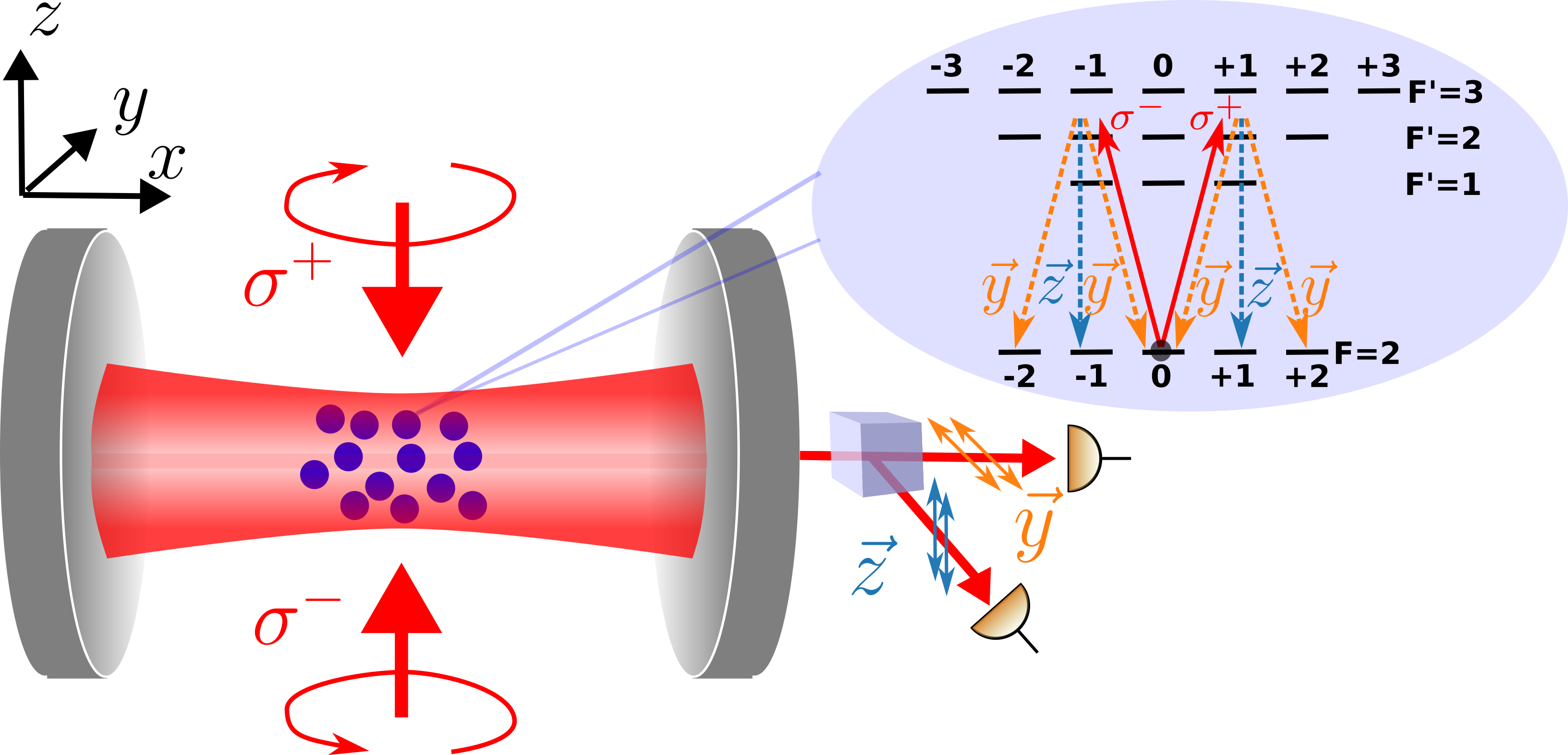}
\caption{Scheme of the experiment with $\sigma^+$-$\sigma^-$ laser drives on the D2 line of ${}^{87}$Rb atoms in the cavity and the polarization-resolved photon detection. The two-photon transitions induced by the coherent drives $\sigma^\pm$ are illustrated in the example for an initial state $m_F=0$. Cavity photons in three possible scattering channels are created: $\Delta m=0 \;$ (Rayleigh), and $\Delta m=\pm 1, \pm2 \;$ (Raman) transitions with respect to the quantization axis $\vec z$, the corresponding cavity mode polarizations are indicated, which are directed to separate detectors.}
\label{fig:Scheme}
\end{figure}

Our cavity QED system consists of a high-finesse optical cavity with a fundamental mode of bare frequency $\omega_c$ and linewidth $\kappa=2 \pi \times 4$ MHz (HWHM), interacting with an ensemble of cold ${}^{87}$Rb atoms prepared in the hyperfine state $F=2$. The maximum atom-cavity coupling constant is $g=2 \pi \times 0.33$ MHz for the cycling transition. The atoms are illuminated from a direction perpendicular to the cavity axis by two counterpropagating laser beams circularly polarized in $\sigma^+ - \sigma^-$ configuration(\cref{fig:Scheme}). The laser drive frequency $\omega$ is tunable in a broad spectral range across the atomic transitions $F=2 \leftrightarrow F'=1,2,3$, meanwhile the cavity resonance $\omega_c$ is frequency locked to the drive at a tunable detuning $\Delta_c\equiv \omega-\omega_c$. For small cavity detuning $|\Delta_c| \lesssim \kappa$, photons are scattered by the atoms into the cavity. This is the only light source for the undriven  cavity.  The field outcoupled from the cavity is directed along polarization-selective ($\vec{y}$, $\vec{z}$) paths to high efficiency single photon counters. 

The outcoupled field modes with linear polarizations $\vec{y}$ and $\vec{z}$ correspond to two degenerate modes of the cavity that are populated by different scattering processes. The counterpropagating $\sigma^+$ and $\sigma^-$ polarized beams form a linearly polarized field with a direction varying cyclically in the ($\vec{x}$, $\vec{y}$) plane along the propagation axis. The component $\vec{x}$ is parallel to the cavity axis, so it is missing from the cavity field. The transverse drive has no $\vec{z}$ polarization component. Therefore, only the $\vec{y}$ polarization leads to coherent Rayleigh scattering into the cavity. Throughout the experiment, the drive laser power was weak to keep the atoms in the low excitation limit where they behave as unsaturated linear scatterers.  Proportionally to the population in the excited states, which scales thus linearly with the drive power, the multiplet structure of the ground state opens Raman scattering channels. In a single absorption-emission cycle, the atomic ground state can change, as depicted in \cref{fig:Scheme}. With respect to the quantization axis $\vec{z}$, the input light induces $\Delta m=\pm 1$ transitions to the excited state manifold. The $\vec{z}$ polarized cavity mode stimulates two-photon Raman processes ending up with $\Delta m=\pm 1$ magnetic sublevel change. There are also $\Delta m=\pm 2$ Raman transitions which populate the $\vec{y}$ polarized mode. This is an incoherent source on top of the coherent Rayleigh scattering, this latter corresponds to the $\Delta m=0$ transition in the low saturation regime. We note that the sublevels were degenerate since compensation coils were used to annul the magnetic field in the vicinity of the atoms. 

The single-atom Hamiltonian consists of the bare and two interaction terms, i.e., the atom-cavity coupling and the laser drive, which are, in the rotating-wave approximation and in a frame rotating at the drive-laser frequency, respectively, 
\begin{multline}
H_0= - \Delta_c\,(a_z^\dagger a_z + a_y^\dagger a_y)
    - \sum_{m',F'} (\omega - \omega_{F'}) \dyad{F', m'}{F', m'}
\\
H_{\text{cav}} = 
 i g \cos(k_c x) \sum_{m,F'} \Bigl[
      a_z^\dag c_{F';m,m} \dyad{F,m}{F',m} 
\\
    + \frac{i}{\sqrt{2}} a_y^\dag \Big( c_{F';m,m-1} \dyad{F,m}{F',m-1} \\ 
    + c_{F';m,m+1} \dyad{F,m}{F',m+1} \Big) 
\Bigr]+ \text{H.c.}  \\
\\
H_{\text{drive}} =  i \eta_+ \sum_{m,F'} e^{-i k z} c_{F';m,m+1} \dyad{F,m}{F',m+1} + \text{H.c.} 
  \\
+ i \eta_- \sum_{m,F'} e^{i k z} c_{F';m,m-1} \dyad{F,m}{F',m-1} + \text{H.c.} 
\label{eq:SingleAtomHamiltonian}
\end{multline}
where $a_y$ and $a_z$ are the field amplitude operators of the cavity modes with $\vec{y}$ and $\vec{z}$ polarizations, respectively, the atomic states $\ket{F,m}$ are defined with respect to the quantization axis $\vec{z}$, $c_{F';m,m'}$ is the Clebsch-Gordan coefficient for the $F=2,m\leftrightarrow F',m'$ transition, $\eta_\pm$ is an effective laser drive amplitude for the $\sigma^\pm$ polarization beams, and $x$ and $z$ are the atomic position coordinates. 

Photon scattering in the weak drive limit can be described perturbatively when both the atom-laser and the atom-cavity interactions are dominated by the detunings and dissipation rates. The scattering amplitude from the initial $\ket{F,m;0}$, where 0 refers to the cavity vacuum, to a final $\ket{F,m+\Delta m ; 1_{y(z)}}$ atomic state with a single photon either in the $\vec{y}$ or $\vec{z}$ polarized mode can be formally expressed as \cite{yan_superradiant_2023,masson_state-insensitive_2024}
\begin{widetext}
\begin{equation}
    S_{m;y(z)}^{\Delta m} = \sum_{F',m'} \frac{\mel{F,m+\Delta m; 1_{y(z)}}{H_{\text{cav}}}{F',m';0} \mel{F',m';0}{H_{\text{drive}}}{F,m;0}}{\omega-\omega_{F'}+i\gamma_{F'}} \, ,
    \label{eq:ScatteringAmplitude}
\end{equation}
\end{widetext}
where $\Delta m \in \qty{0,\pm1,\pm2}$. The transition matrix elements are evaluated between eigenstates of the bare atomic Hamiltonian \cite{SM}. All the allowed intermediate excited hyperfine states $F'$ and sublevels $m'$ have to be taken into account, which brings in quantum interference at the single atom level. The two-photon scattering rate from an initial magnetic sublevel $m$ into the  $\vec{y}$ or $\vec{z}$ polarized cavity modes is equal to $W_{m;y(z)}=\sum_{\Delta m}\abs{S_{m;y(z)}^{\Delta m}}^2$, where the sum over $\Delta m$ accounts for all possible final states.  

The scattering amplitude above refers to a single atom. When many atoms are in the cavity, there is an interference between the Rayleigh scattered components from different atoms, i.e., the complex amplitudes for $\Delta m=0$ processes should be added up. These amplitudes depend on the atomic positions within the mode function $\cos{kx}$. As it was demonstrated recently \cite{gabor_demonstration_2025}, the quasi-homogeneous distribution of the cloud leads to destructive spatial interference, and only the density fluctuations lead to an incoherent scattering. The scattered intensity is thus proportional to the atom number and can be described by summing the scattering rates. For the Raman scattered part ($\Delta m\neq0$), obviously, there is no interference since the change in the atomic sublevel stores `which way' information about the scattered photon. Both for the Raman and Rayleigh scattering, the distribution in the Zeeman sublevel manifold has to be taken into account by an appropriate averaging of the rates.  

\section{Results}

\begin{figure}[htbp]
  \centering
    \includegraphics[width=0.89\linewidth]{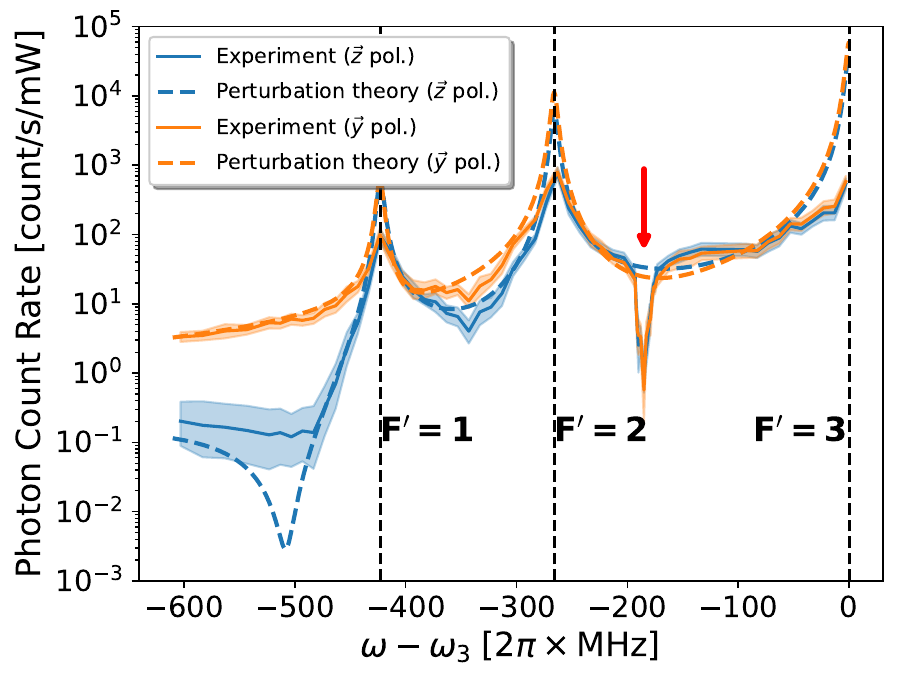}    
  \caption{Polarization-resolved photon count rates as a function of the drive detuning.  The measurement data (solid lines with shaded area representing the width of the distribution at half maximum) are compared with the single-atom scattering rate model (dashed line).  Vertical dashed lines indicate the transition resonances to excited hyperfine states. The arrow indicates the collective inhibition of scattering at the magic frequency. }
  \label{fig:ExperimentVSTransitionRate}
\end{figure}

The polarization resolved photon scattering rate into the cavity has been measured in a wide range of frequencies covering the optical transitions to the excited hyperfine states. The cavity was locked to resonance with the laser drive frequency ($\omega_c=\omega$) to maximize photon collection in the cavity. The average and the standard deviation of the data of 100 repeated measurements for each atomic detuning is shown in \cref{fig:ExperimentVSTransitionRate} together with a fit based on the perturbative expression from \cref{eq:ScatteringAmplitude}. A single multiplicative fitting parameter representing the combined effect of the drive power and the effective atom number was used, which was determined in the far red detuning limit on the $\vec{y}$ component (orange curve). With this parameter, the perturbative theory gives an overall good account for the large features of the experimental observations for both polarizations, including the $\vec{z}$ component (blue curve). For example, the measurement confirmed the prediction of suppressed Raman scattering at the detuning of -506 MHz \cite{yan_superradiant_2023}. The $\vec{z}$-polarized signal arising from $\Delta m=\pm1$ scattering channels nearly vanishes, in agreement with the dip of the theoretical curve. The higher experimental data can be explained by dark counts and background light scattering in the $\vec{z}$-polarization channel. The suppression of the Raman scattering marks a destructive interference among the transition paths via the hyperfine states $F' = 1,2,3$ irrespective of the magnetic sublevel $m$. It is interesting that the drive frequency is red-detuned with respect to all $F'$ excited states, so destructive interference within a single atom can be obtained only because of the different signs of Clebsch–Gordan coefficients. Similar interference insensitive to the magnetic sublevel at a magic frequency has been reported in the absorption of a linearly polarized light beam by a vapor of alkali atoms \cite{givon_magic_2013}.

Deviations of the measured data from single-atom scattering theory reveal the presence of collective effects. The most prominent feature is a pronounced dip at a detuning of $-185$ MHz, observed for both polarizations. In the vicinity of this dip, the measured scattering rate exhibits a marked departure from the $1/\Delta$ dependence expected far from resonance in the perturbative regime, indicated by the dashed lines in \cref{fig:ExperimentVSTransitionRate}. We found that the coincidence of the dips in the $\vec{y}$ and $\vec{z}$ polarizations cannot be reproduced from the single-atom scattering model whatever the population  distribution is in the ground-state sublevels. In this case, the perturbative expression fails because the unperturbed bare states are taken as intermediate excited states, whereas the atoms cannot be excited individually in the strong collective coupling regime $\sqrt{N} g > \kappa, \gamma$.   

\begin{figure}[htbp]
\centering
\includegraphics[width=0.89\linewidth]{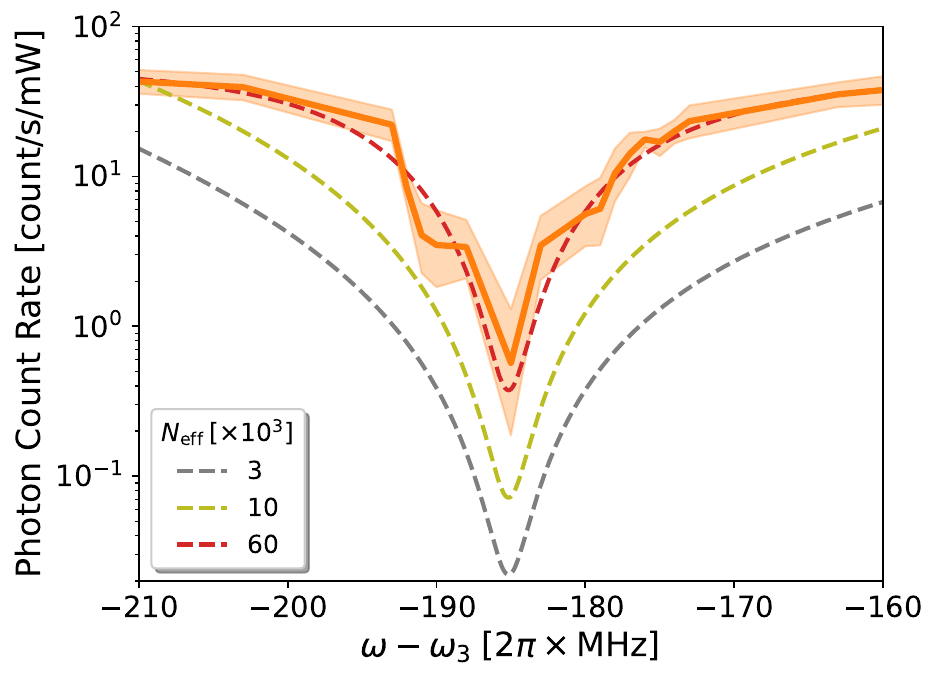}
\caption{Photon scattering rate into the $\vec{y}$-polarized mode near the dip at $-185$ MHz. The average of measured data are plotted by the orange curve, the shaded area representing the standard deviation of 100 measurements. Dashed lines give the polariton theory curves for different effective atom numbers $N_{\text{eff}}$ indicated in the legend. These curves are plotted with a frequency offset to match the experimental dip position.  The cavity detuning was $\Delta_c=0$.}
\label{fig:DIP_geff}
\end{figure}
Although single atoms have only a small effect on the cavity field, they add up to a significant collective effect. The cavity field has, on the other hand, a back-action on each of the atoms.  In such an interacting system the normal modes are the independent degrees of freedom. These are the so-called \emph{polaritons} which mix the cavity mode amplitudes and the atomic dipoles, separately for both polarizations $\vec{y}$ and $\vec{z}$. In the strong collective coupling regime, cavity photons can be created only in the form of polariton excitations. However, the coupling of the polaritons to the laser drive is complex because of the many atoms in different positions and Zeeman sublevels. The coherent polariton excitations generated by the coherent laser drive can be treated within a simple cavity QED mean-field model \cite{SM}. Coherent field is generated only in the $\vec{y}$ polarized mode, the corresponding intensity is
\begin{equation}
    \abs{a_y}^2=\eta^2 g^2 \mathcal{S}\frac{\abs{\mathcal{P}(\omega)}^2}{\abs{i\Delta_c - \kappa+g^2N_{\text{eff}}\mathcal{P}(\omega)}^2} \, ,
    \label{eq:phnumCoh}
\end{equation}
where the $\vec{y}$-polarized laser field has an effective amplitude $\eta$,  $\mathcal{S}=\abs{\sum_{a=1}^N\cos(k_cx_a)\cos(kz_a)}^2$ accounts for the positions of the scatterers within the cavity mode and the transverse driving field, and $N_{\text{eff}} = \sum_{a=1}^N\cos^2(k_cx_a)\approx N/2$ is the effective atom number. The denominator includes the Lorentzian resonance of the cavity with a collective displacement $g^2N_{\text{eff}}\mathcal{P}(\omega)$ due to the atoms. The effect of the atoms is expressed by the collective polarizability function
\begin{equation}
    \mathcal{P}(\omega)= -i\sum_{F'}\frac{\expval{c_{F'}^2}}{\omega-\omega_{F'} + i\gamma_{F'}} \, .
\end{equation}
Summation over the possible excited hyperfine states leads again to interference. However, in the collective polarizability of a many atom ensemble, the Clebsch-Gordan coefficients have to be averaged over the population distribution in the magnetic sublevels \cite{SM}. There is a "magic" frequency $\omega^*$ between the resonances to $F'=2$ and $F'=3$ for which the polarizability is strongly suppressed.  At this frequency $\omega^*$ of the drive, the excitation of the collective $\vec{y}$ polariton is blocked, represented by the vanishing numerator in \cref{eq:phnumCoh}. Perfect cancellation of $ \mathcal{P}$ could take place if the imaginary part $\gamma_{F'}$ were absent. In fact, the atoms entirely decouple from the $\vec{y}$-polarized mode at $\omega^*$, in accordance with the denominator reducing to that of an empty cavity. Since there is no other excitation in the system but the coherent $\vec{y}$ polarized drive, the excited state population is eliminated for the entire ensemble of atoms. No spontaneous decay processes resulting in either $\vec{y}$ or $\vec{z}$ polarized cavity photons occurs then. The collective inhibition of Rayleigh scattering entails a lack of Raman scattering from individual atoms and the dips in the $\vec{y}$ and $\vec{z}$ polarized channels overlap. 

The vanishing of $\mathcal{P}$ at a magic frequency is determined by internal atomic properties, i.e., the Clebsch–Gordan coefficients, and the population distribution in the magnetic sublevels of the ground state. For uniform distribution, the theoretical value is $-192$ MHz whereas we observed the dip at about $\Delta^*= -185$ MHz. We attribute this difference to a small deviation from the uniform distribution in the sublevels.  

The shape of the dips is in good qualitative agreement with \cref{eq:phnumCoh} from the polariton model. By zooming in on the vicinity of the magic frequency, \cref{fig:DIP_geff} shows the measured photon rate together with theory curves with two fitting parameters: one for the multiplicative prefactor and one for the effective atom number $N_\text{eff}$. The larger the atom number, the narrower the region of collective inhibition of scattering. The experimental data are well reproduced with $N_\text{eff}= 6\times 10^4$ which is within a factor of 4 of its expected value.  The polarizability can be expanded around the minimum at the magic frequency $\omega^*$ to leading order as $\mathcal{P}(\omega) \approx -i\,p_1 \delta $ with the small variable $\delta \equiv \omega-\omega^*$ and some linear coefficient $p_1$.
The coherent intracavity photon number is then $\abs{a_y}^2 \propto \delta^2 / (\kappa_{\text{eff}}^2 + \delta^2 )$, with $\kappa_\text{eff}= \kappa/(g^2 N_{\text{eff}} p_1)$, when the transverse drive is tuned in resonance with the cavity ($\Delta_c=0$).  This approximation describes the quadratic shape of the dip  within the range $\delta< \kappa_{\text{eff}}$. The saturation to constant away from the minimum $\delta > \kappa_{\text{eff}}$, roughly at a detuning of 15 MHz from the magic frequency, is in agreement with the plateau exhibited in the larger range shown in \cref{fig:ExperimentVSTransitionRate}.  

\begin{figure}[htbp]
\centering
\includegraphics[width=\linewidth]{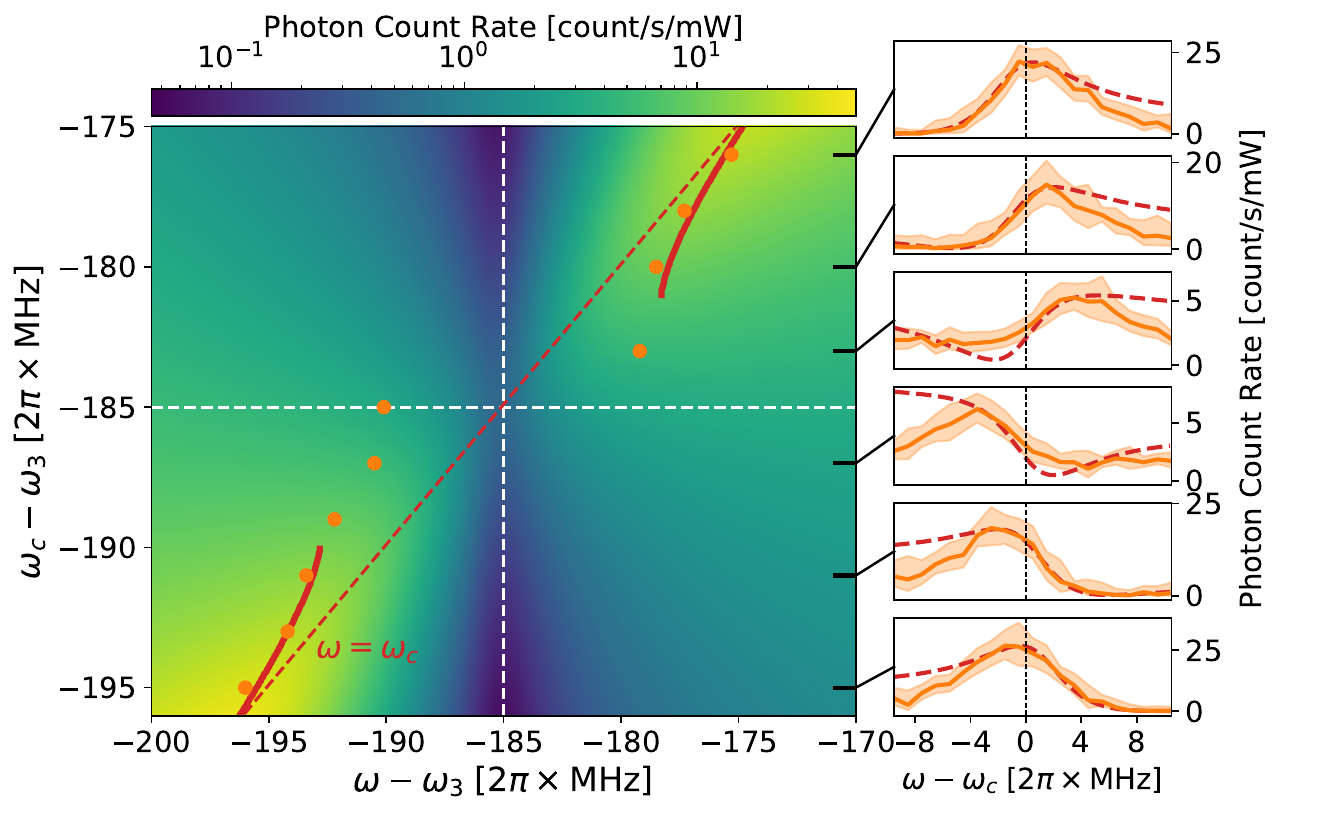}
\caption{Polariton-mediated coherent photon scattering rate into the $\vec{y}$-polarized cavity mode as a function of the drive and cavity frequencies near the interference dip at –185 MHz. 2D color-coded plot presents the steady-state solution of \cref{eq:phnumCoh}, the two sections of thick red lines indicate the maxima as a function of the drive $\omega-\omega_3$. This was fit to the experimentally measured  maxima of the cavity photon number for selected cavity frequencies, given by orange points. For some of them, the drive frequency dependence is shown (orange line with shaded area representing the standard deviation) in the right column of panels. Red dashed lines in these panels represent the theory curves, for which the maxima are not well defined between –181 MHz and –190 MHz.}
\label{fig:omega-omegacMAP}
\end{figure}
The laser drive excites the collective polaritons to an extent depending on the cavity detuning $\Delta_c$ in accordance with \cref{eq:phnumCoh}. The scattering rate as a function of the atomic and cavity detunings is mapped in a 2D colour plot in  \cref{fig:omega-omegacMAP}.  Further away from the magic frequency on both sides, the almost bare Lorentzian of the cavity mode with $\kappa \sim 4$ MHz linewidth can be seen along vertical cuts (fixed drive and atomic frequency), and the peak falls on the diagonal red dashed  line representing $\omega=\omega_c$. This simple landscape is cut through vertically by the dark valley at $\omega-\omega_3=-185$ MHz appearing due to the collective suppression of scattering into the resonator. A dip is imprinted into the Lorentzian, which leads to a maximum (orange dots) deviating from the $\omega=\omega_c$ line.  The two red curves in \cref{fig:omega-omegacMAP} trace the maxima of the spectrum given in \cref{eq:phnumCoh}. Their position depends strongly on both the cavity linewidth and the collective coupling strength: the former enhances, while the latter reduces the distance from the $\omega=\omega_c$ resonance line. The experimental result is reproduced by $N_{\text{eff}}\approx 3 \times 10^4$. This value is half of the one obtained from fitting to the shape of the dip in \cref{fig:DIP_geff}. This discrepancy can be attributed to the distortion of the dip shape by the run-to-run fluctuations due to the imperfect control of the atom number in the sample.

Finally, we note that the strong collective coupling between the atomic ensemble and the cavity mode also accounts for the deviation of the experimental data from single-atom scattering theory at the hyperfine resonances, indicated by the dashed lines in \cref{fig:ExperimentVSTransitionRate}. This behaviour is related to the well-known normal-mode splitting \cite{thompson_observation_1992}. As a consequence of the collective vacuum Rabi splitting effect, the maxima in the spectrum of the coupled system are shifted by $\pm \sqrt{N} g$ from the origin \cite{gabor_demonstration_2025}. Accordingly, the rate of scattering is reduced at exact resonance $\omega=\omega_{1,2,3}$, whereas the perturbative expression retains the bare frequencies in the denominator of the scattering rate in  \cref{eq:ScatteringAmplitude}.     

\section{Methods}

Cold ${}^{87}$Rb atoms are collected in a magneto-optical trap and cooled down to 10 $\mu$K using polarization-gradient cooling. The atoms are optically pumped into the low-field-seeking $(F,m_F)=(2,2)$ state and loaded into a magnetic quadrupole trap, which is adiabatically translated into the mode volume of a high-finesse optical cavity. After switching off the magnetic field, two transverse pump beams are applied, and their frequency is scanned across the $F=2 \leftrightarrow F'={1,2,3}$ manifold in 10 MHz steps, and in 1 MHz steps in the vicinity of the dip. At each frequency point, the pump power is adjusted to remain below the single-atom saturation threshold. A repumper addressing the $F=1 \leftrightarrow F'=2$ transition is applied simultaneously to maintain population in the cycling manifold. For each pump-frequency setting, 100 experimental repetitions are performed.
The cavity output is analyzed in orthogonal linear polarizations using a polarizing beam splitter, and each polarization component is detected with a single-photon counting module. The overall detection efficiency is approximately 60 \%, including both fiber-coupling losses and detector quantum efficiency. Photon counts are integrated over the first 1 ms of illumination with a temporal resolution of 1 $\mu$s.
\section{Discussion}

One of the main reasons for using optical resonators in a variety of atom-light systems is that the commonly coupled resonator mode accumulates the effect of the atoms, thereby creating the possibility of many-atom enhancement in sensitivity. Of particular interest are the schemes in which the individual atomic properties, such as position, Zeeman state, or other inhomogeneities average out and a global action stands out in the cavity response. Well-known examples are (i) the normal-mode splitting for closely resonant atoms and cavity mode \cite{thompson_observation_1992,tuchman_normal-mode_2006,courteille_photonic_2025}, (ii) and the collective dispersive shift \cite{elsasser2004optical,dombi2021collective} for large atomic detuning. These effects can be measured in the transmission of a weak probe light through the cavity. A new strong collective coupling effect has been discovered and presented in this paper, which appears in the light scattering of laser-driven atomic ensemble into the cavity. It takes place in the large atom-cavity detuning limit, however, at a well-defined magic frequency which is defined by a quantum interference between the excitations of different atomic hyperfine states. It is also because of the quantum interference that the spectral feature obtained near the magic frequency is significant: we observed two orders of magnitude variation in cavity photon number. Note that the atomic detuning at the magic frequency would lead to only a very small collective dispersive shift, well below the mode linewidth ($\sim \kappa/4$). The magic frequency is defined by the ensemble of atoms, averaging over their internal states, which gives a robustness to the dip signal. Further improvement can be obtained by reducing the run-to-run noise by means of technical developments, and then the effect can be the basis of precise atomic reference.

\section{Acknowledgments}
This research was supported by the Hungarian National Research, Development and Innovation Office (Grant Nos. 2022-2.1.1-NL-2022-00004 and 2025-3.1.1-ED-2025-00011), the ERANET COFUND QuantERA programme (MOCA 2019-2.1.7-ERA\_NET-2022-00041), the QuantERA II Programme (V-mag 2024-1.2.2-ERA\_NET-2024-00012), and by the second Swiss Contribution MAPS (Grant No. 230870). AD and TWC acknowledges support from the János Bolyai research scholarship of the Hungarian Academy of Sciences.

\bibliography{MagicWavelength}

\end{document}